\documentclass[twocolumn,floats,showpacs,prl,amssymb,floatfix,nofootinbib,balancelastpage,superscriptaddress,amsmath]{revtex4}
\usepackage{epsfig}
\usepackage{graphicx}
\usepackage{dcolumn}
\usepackage{bm}
\usepackage{amssymb}
\usepackage{amsmath}

                              \newlength{\strikewidth}
                              \newlength{\strikelength}
\begin{document}

\title{Solar System tests {\it do} rule out $1/R$ gravity}

\author{Adrienne L. Erickcek, Tristan L. Smith, and Marc
Kamionkowski}
\affiliation{California Institute of Technology, Mail Code
130-33, Pasadena, CA 91125}


\begin{abstract}
Shortly after the addition of a $1/R$ term to the Einstein-Hilbert
action was proposed as a solution to the cosmic-acceleration
puzzle, Chiba showed that such a theory violates
Solar System tests of gravity.  A flurry of recent
papers have called Chiba's result into question.  They argue
that the spherically-symmetric vacuum
spacetime in this theory is the
Schwarzschild-de Sitter solution, making this theory consistent with Solar System tests.  
We point out that
although the Schwarzschild-de Sitter solution exists in this
theory, it is not the {\it unique} spherically-symmetric
vacuum solution, and it is {\it not} the solution that
describes the spacetime in the Solar System.  The solution that correctly
matches onto the stellar-interior solution differs from
Schwarzschild-de Sitter in a way consistent with Chiba's claims.
Thus, $1/R$ gravity is ruled out by Solar System tests.
\end{abstract}




\maketitle

The discovery of an accelerated cosmic expansion
\cite{Perlmutter:1998np,Riess:1998cb}  has led to a
flurry of theoretical activity.  One class of solutions to the
cosmic-acceleration puzzle consists of modifications to the
general-relativistic theory of gravity.  One particular proposal
is the addition of a $1/R$ term to the Einstein-Hilbert action
\cite{Carroll:2003wy,Capozziello:2003tk}.
Such a term gives rise to a vacuum solution with constant
curvature, the de Sitter spacetime, rather than the Minkowski vacuum
of the usual Einstein-Hilbert action.

Shortly after this proposal, Chiba \cite{Chiba:2003ir} argued
that this theory is
inconsistent with Solar System tests of gravity.  In particular,
he showed that the theory is equivalent to a scalar-tensor
theory that is known to make Solar System predictions that
conflict with measurements.  

Since then, however, there have
been a number of papers arguing or implying that Chiba's analysis is flawed
\cite{Rajaraman:2003st,Multamaki:2006zb,Faraoni:2006hx,Ruggiero:2006qv,Allemandi:2005tg}.
The crux of the counter-argument is that $1/R$ theories admit as a static
spherically-symmetric solution the usual vacuum
Schwarzschild-de Sitter spacetime.  Apart from a cosmological
constant that is too small by many orders of magnitude to affect
anything observable in the Solar System, these
solutions are just the usual Schwarzschild solution.
Consequently, they argue, there is no effective difference
between the Solar System spacetime in these models and that in
ordinary general relativity.

Here we point out that these arguments are incorrect, and that
Chiba was right.  The crucial point is that although the
Schwarzschild-de Sitter spacetime is indeed a spherically-symmetric vacuum solution to
the $1/R$ equations of motion, it is not the {\it unique} spherically-symmetric vacuum solution in this theory.  The
correct solution is determined by matching onto the solution in
the interior of the star.  When this is done correctly, it is
found that the Schwarzschild-de Sitter spacetime does {\it not}
describe the spacetime around the
Sun, and that Chiba's result stands.  This misunderstanding has
now propagated through a number of papers.  There are moreover a number of other papers
that cite these incorrect papers in a way that suggests that
they may be onto something.  We thought it worthwhile to correct
the error before it propagates any further.

Before describing the correct spherically-symmetric spacetime
for $1/R$ gravity, we consider a very simple and
analogous problem that illustrates what is going on.  
Suppose we wanted to know the electric field around a
spherically-symmetric charge distribution $\rho(r)$ confined to
radii $r<R$.  For radii $r>R$,
the Poisson equation $\nabla^2 \phi = 4 \pi \rho$ relating the
electric potential $\phi$ to the charge-density distribution
$\rho$ reduces to $\nabla^2 \phi =0$.  A spherically symmetric
solution to this equation, one might argue, is $\phi=0$,
implying no electric field.  This is clearly incorrect.

What went wrong?  Although $\phi=0$ is indeed a spherically
symmetric solution to $\nabla^2 \phi =0$, it is not the {\it
unique} solution.  Another solution is $\phi = c/r$, for $r>R$.  The
constant $c$ in this equation is furthermore fixed in this case
to be $c=Q$, where $Q=\int \rho \ \mathrm{d}^3x$ is the total charge, by
integrating the right- and left-hand sides of the Poisson 
equation $\nabla^2 \phi =4\pi \rho$ over the entire volume.

In brief, something similar happens in $1/R$ gravity.  The
differential equations for the metric components $g_{tt}(r)$ and
$g_{rr}(r)$ 
are supplemented by a differential equation for the curvature
$R$, as we will see below.  The three differential equations
have the Schwarzschild-de Sitter spacetime as a solution, but
these vacuum solutions do not match onto the solutions in the
presence of a source (i.e., the Sun).  There is an additional
vacuum solution that correctly
matches onto the solution in the presence of the source.

Now the details: the gravitational action of $1/R$ gravity,
\begin{equation}
     S=\frac{1}{16 \pi G} \int\, \mathrm{d}^4x\, \sqrt{-g} \left( R -\frac{\mu^4}{R} \right) + \int\, \mathrm{d}^4x\, \sqrt{-g} 
     {\cal L}_{\rm M},
\label{action}
\end{equation}
may be varied with respect to the metric $g_{\mu\nu}$ to obtain
the field equation  \cite{Carroll:2003wy}
\begin{eqnarray}
8\pi G T_{\mu\nu}&=& \left({1+\frac{\mu^4}{R^2}}\right)R_{\mu\nu} 
 - \frac{1}{2}\left({1-\frac{\mu^4}{R^2}}\right)Rg_{\mu\nu}\nonumber\\
&&+\mu^4 \left(g_{\mu \nu} \nabla_{\alpha}\nabla^{\alpha} - \nabla_{\mu}\nabla_{\nu}\right) R^{-2}.
\label{fieldeq}
\end{eqnarray}
We begin by using the trace of the field equation to determine
the Ricci scalar $R$.  Contracting Eq.~(\ref{fieldeq}) with the
inverse metric yields
\begin{equation}
     \Box  \frac{\mu^4}{R^2} -  \frac{R}{3} + \frac{\mu^4}{R} =
     \frac{8 \pi G T}{3} , 
\label{trace}
\end{equation}
where  $T\equiv g^{\mu\nu}T_{\mu\nu}$.  

The constant-curvature vacuum solution is obtained by setting
$T=0$ and  $\nabla_\mu R=0$.  It is $R^2 = 3 \mu^4$,
corresponding to the de Sitter spacetime with Hubble parameter
$H^2 = \mu^2/(4\sqrt{3})$, equivalent to the
general-relativistic vacuum solution with a cosmological
constant $\Lambda=3H^2 = \sqrt{3} \mu^2/4$.  The metric for this
spacetime can be written as a static spherically-symmetric
spacetime:
\begin{equation}
     ds^2 = -\left(1- H^2 r^2\right) dt^2 +
     \left(1-H^2 r^2\right)^{-1} dr^2 + r^2
     d\Omega^2.
\label{dSmetric}
\end{equation}
To match the observed acceleration of the universe, the
effective cosmological constant must be set to $\Lambda \sim
\mu^2 \sim H^2 \sim 10^{-56}$ cm$^{-2}$.

We now consider the spacetime in the Solar System in this theory.
First of all, the distances ($\sim10^{13}$ cm) in the Solar System
are tiny compared with the distance $\mu^{-1}\sim 10^{28}$ cm, so
$\mu r \ll1$ everywhere in the Solar System.  Moreover, the
densities and velocities in the Solar System are sufficiently
small that we can treat the spacetime as a small perturbation to
the de Sitter spacetime.  The spacetime should also be
spherically symmetric and static.  The most general static
spherically-symmetric perturbation to the vacuum de Sitter spacetime given by Eq.~(\ref{dSmetric})
can be written
\begin{eqnarray}
     ds^2 &=& -\left[1+a(r)- H^2 r^2\right] dt^2 \nonumber\\
     &&+ \left[1+b(r)-H^2 r^2\right]^{-1}dr^2 + r^2 d\Omega^2,
\label{eqn:ourmetric}
\end{eqnarray}
where the metric-perturbation variables $a(r),b(r) \ll1$.  In
the following, we work to linear order in $a$ and $b$, and also
recall that $\mu r\ll1$.  However, $a,b$ are {\it not} necessarily
small compared with $\mu r$.

We now return to the trace of the field equation, given by
Eq.~(\ref{trace}), and solve it for the Ricci scalar $R(r)$ in
the presence of the Sun.   We write the trace equation in terms
of a new function,
\begin{equation}
c(r) \equiv -\frac{1}{3}+\frac{\mu^4}{R^2(r)},
\label{defc}
\end{equation}
and demand that $c(r)\rightarrow 0$ as $r\rightarrow \infty$ so
that $R$ approaches its background value of $\sqrt{3}\mu^2$ far
from the source of the perturbation.  Therefore, $c(r)$
parameterizes the departure of $R$ from the vacuum solution, and
we anticipate that $c(r)$ will be the same order in the
perturbation amplitude as the metric perturbations $a(r)$ and
$b(r)$.  In terms of $c(r)$, Eq.~(\ref{trace})
becomes an {\it exact} equation,
\begin{equation} 
     \Box c(r) + \frac{\mu^2 c}{\sqrt{c+\frac{1}{3}}} = \frac{8 \pi G}{3}T.
\label{boxc}
\end{equation}
In the Newtonian limit appropriate for the Solar System, the
pressure $p$ is negligible compared to the energy
density $\rho$, and so $T=-\rho$.  Neglecting terms that are
higher order in $a(r)$, $b(r)$, and $\mu^2 r^2$, we are able to
rewrite Eq.~(\ref{boxc}) as
\begin{equation}
     \nabla^2 c + \sqrt{3} \mu^2 c = - \frac{8 \pi G}{3} \rho, 
\label{linearized}
\end{equation}
where $\nabla^2$ is the flat-space Laplacian operator.  Note
that in writing this equation, which is linear in $c(r)$, we
have also neglected higher-order terms in $c(r)$.  Below, we
will check that the solutions we obtain have $c(r) \ll1$
everywhere, consistent with our assumptions.  The Green's
function for Eq.~(\ref{linearized}) is $-\cos(3^{1/4}\mu r)/(4\pi r)$.
Convolving this with the density gives us the solution to
Eq.~(\ref{linearized}).  However, we are restricting
our attention to the region where $\mu r \ll 1$, so the Green's
function reduces to that for the Laplacian operator.  Therefore
the equation we need to solve is $\nabla^2 c = - (8 \pi G \rho)/3$.
Integrating the right-hand side over a spherical volume of
radius $r$ gives us $-8 \pi G m(r)/3$, where $m(r)$ is the mass
enclosed by a radius $r$.  Using Gauss's law to integrate the
left-hand side gives us $4\pi r^2 c^{\prime}(r)$, where the prime denotes differentiation with 
respect to $r$.  Thus, the equation
for $c(r)$ becomes
\begin{equation}
     \frac{\mathrm{d}c}{\mathrm{d}r} =-\frac{2G m(r)}{3 r^2}
     \left[1 + {\cal O}(\mu r)\right].
\label{csol}
\end{equation}
Integrating Eq.~(\ref{csol}) and using the boundary condition
that $c \rightarrow 0$ as $r\rightarrow\infty$ gives us the
solution $c(r) = (2/3)(GM/r)[1+{\cal O}(\mu r)]$ for
$r>R_\odot$.  Note also that integration of the equation for
$c^\prime(r)$ to radii $r<R_\odot$ inside the star implies that
the scalar curvature $R$ remains of order $\mu^2$, even inside
the star.  We thus see that $c \ll 1$, so we were justified in
using the linearized equation for $c(r)$.  

This solution for $c(r)$ implies that
\begin{equation}
     R=\sqrt{3} \mu^2 \left( 1 - \frac{GM}{r} \right), \qquad
     r>R_\odot.
 \label{extR}
\end{equation}
We have thus shown that $R$ is not constant outside the star and
have already arrived at a result at odds with the
constant-curvature Schwarzschild-de Sitter solution.  Notice that
had we ({\it incorrectly}) used $\rho=0$ in
Eq.~(\ref{linearized}), then the equations would have admitted the
solution $c(r)=0$; i.e. the constant-curvature solution.  However, {\it this would be
incorrect, because even though} $\rho=0$ {\it at} $r>R_\odot$,
{\it the solution to the differential equation at $r>R_\odot$
depends on the mass distribution $\rho(r)$ at $r<R_\odot$}.  In
other words, although the Schwarzschild-de Sitter solution is a
static spherically-symmetric solution to the vacuum Einstein
equations, {\it it is not the solution that correctly matches onto
the solution inside the star}.  Note further that the solution
for $R$ both inside and outside the star is (to linear order in
$c$), 
\begin{equation}
     R=\sqrt{3} \mu^2 \left[1 -  \frac{3}{2}c(r)\right ].
\end{equation}
Clearly, $1/R$ gravity produces a spacetime inside the star that is {\it
very} different from general relativity.  This result shows that in this theory one should not assume that $R=8 \pi G\rho$; 
this has lead to some confusion \cite{Shao:2005wt,Cembranos:2005fi,Dolgov:2003px}.

To proceed to the solutions for $a(r)$ and $b(r)$, we rearrange
the field equation for $1/R$ gravity [Eq.~(\ref{fieldeq})] to
obtain equations,
\begin{eqnarray}
     R_{\mu\nu} = &&
     \left({1+\frac{\mu^4}{R^2}}\right)^{-1}\left[{8\pi G T_{\mu
     \nu}+\frac{1}{2}\left({1-\frac{\mu^4}{R^2}}\right)Rg_{\mu\nu}
     }\right.\nonumber\\ 
     &&{ -\mu^4 \left({g_{\mu \nu}
     \nabla_{\alpha}\nabla^{\alpha} -
     \nabla_{\mu}\nabla_{\nu}}\right) R^{-2} } \bigg],
\label{RicciTensor}
\end{eqnarray}
for the Ricci tensor in terms of the Ricci scalar.
When the expression for $R$ obtained from the trace equation is
inserted into the right-hand side, we obtain equations for the
nonzero components of the Ricci tensor,
\begin{eqnarray}
     R^t_t &=& 3H^2  - 6\pi G \rho -
     \frac{3}{4} \nabla^2 c, \label{Rtt} \\
     R^r_r &=& 3H^2
     -\frac{3c^{\prime}(r)}{2r}, \label{Rrr} \\
     R^\theta_\theta = R^\phi_\phi &=& 3H^2
     - \frac{3}{4}\left(\frac{c^{\prime}(r)}{r} + c^{\prime
     \prime}(r)\right) \label{Rthetatheta},
\end{eqnarray}
where we have neglected terms of order $\mu^2 c$, $G \rho c$ and $c^2$ in all three
expressions.

For the perturbed metric given by Eq.~(\ref{eqn:ourmetric}), the
$tt$ component of the Ricci tensor is (to linear order in small
quantities) $R^t_{t}=3H^2 - (1/2) \nabla^2 a(r)$.
Applying $\nabla^2 c = -(8 \pi G \rho)/3$ to Eq.~(\ref{Rtt}) leaves us
with an equation for $a(r)$,
\begin{equation}
     \frac{1}{2} \nabla^2 a = 4 \pi G \rho,
\end{equation}
plus terms that are higher order in $GM/r$ and $\mu r$.  The solution to this equation parallels that 
for $c(r)$; it is
\begin{equation}
        \frac{\mathrm{d}a}{\mathrm{d}r} = 2 G \frac{m(r)}{r^2}
        \label{aprime}
\end{equation}
both inside and outside the star.  Outside the star, this
expression may be integrated, subject to the boundary condition
$a(r)\rightarrow0$ as $r\rightarrow \infty$, to obtain the
metric perturbation,
\begin{equation}
     a(r) = -\frac{2 GM}{r}, \qquad r>R_\odot,
\end{equation}
exterior to the star.  Note that this recovers the Newtonian
limit for the motion of nonrelativistic bodies in the Solar
System, as it should.

The $rr$ component of the Ricci tensor is (to linear order in
small quantities) $R^r_{r}= 3H^2-(b'/r)-(a''/2)$.
Given our solution for $a^\prime(r)$ and $c^\prime(r) = -(2/3)G m(r) / r^2$,
Eq.~(\ref{Rrr}) becomes a simple differential equation for
$b(r)$,
\begin{eqnarray}
      \frac{\mathrm{d}b}{\mathrm{d}r} &=& \frac{G m(r)}{r^2} -
      \frac{G m^\prime(r)}{r} \nonumber \\ 
      &=& \frac{\mathrm{d}}{\mathrm{d}r}\left[\frac{-G m(r)}{r}\right].
\end{eqnarray}  
Integrating this equation subject to the boundary condition
$b(r)\rightarrow0$ as $r\rightarrow \infty$ gives an expression
for $b(r)$ that is applicable both inside and outside the star:
\begin{equation}
     b(r) = -\frac{G m(r)}{r}.
\end{equation}
This expression for $b(r)$ and Eq.~(\ref{aprime}) for
$a^\prime(r)$ also satisfy Eq.~(\ref{Rthetatheta}) for the
angular components of the Ricci tensor.  The
Ricci scalar [Eq.~(\ref{extR})] is recovered from the Ricci
tensor components if terms higher order in $\mathcal{O}(\mu r^2
G M/r)$ are included in our expressions for $a(r)$ and $b(r)$.

The linearized metric outside the star thus becomes
\begin{eqnarray}
     ds^2 &=& -\left(1- \frac{2GM}{r}-H^2 r^2\right) dt^2 \\
     & &+
     \left(1+ \frac{GM}{r} + H^2 r^2\right) dr^2 + r^2
     d\Omega^2.
\end{eqnarray}
Noting that in the Solar System, $H r \ll 1$ and that the PPN
parameter $\gamma$ is defined by the metric,
\begin{equation}
     ds^2 = -\left(1 - \frac{2GM}{r}\right) dt^2 +
     \left(1 + \frac{2\gamma GM}{r} \right) dr^2 + r^2
     d\Omega^2,
\end{equation}
we find that $\gamma=1/2$ for $1/R$ gravity, in
agreement with Chiba's claims \cite{Chiba:2003ir,Olmo:2005jd}, and prior
calculations; e.g., Refs.~\cite{Brans:1961sx,Will:1993ns}.  We 
note that recent measurements give $\gamma = 1 + (2.1 \pm 2.3)
\times 10^{-5}$ \cite{Bertotti:2003rm, shapiro}.

Other authors have noted that Birkhoff's theorem---that the unique
static spherically-symmetric vacuum spacetime in general
relativity is the Schwarzschild spacetime---is lost in
$1/R$ gravity, and that there may be several
spherically-symmetric vacuum spacetimes.  Although this is true,
what we have shown here is that the Solar System spacetime is
determined uniquely by matching the exterior vacuum solution to
the interior solution.  When this is done correctly, it is found
that the theory predicts a PPN parameter $\gamma=1/2$ in gross
violation of the measurements, which require $\gamma$ to be
extremely close to unity.

A few final comments:  It is important to note that the
structure of $1/R$ gravity (for example, the way matter sources the metric)
is completely different than the structure of general
relativity, even in the limit $\mu \rightarrow0$.  In
particular, the theory does not reduce to general relativity in
the $\mu\rightarrow0$ limit, and this can lead to confusion.
This is due to the fact that the introduction of additional
terms in the Einstein-Hilbert action brings to life a scalar
degree of freedom that lies dormant in general relativity.  
We also note that Chiba's mapping of $f(R)$ theories to
scalar-tensor theories is perfectly valid; it amounts to no more
than a variable change, from $R$ to $\phi \equiv 1+\mu^4/R^2$.  The
trace equation, Eq.~(\ref{trace}), is then equivalent to the
scalar-field equation of motion in the scalar-tensor theory.
Also, the fact that general relativity is not recovered in the
$\mu\rightarrow0$ limit becomes particularly apparent in the
scalar-tensor theory, as we will discuss elsewhere.  Although we
have restricted our analysis, for clarity, to $1/R$ theory,
similar results can also be derived for other $f(R)$ theories.
For example, the correct matching of the exterior and interior
solutions can be used to distinguish between the
spherically-symmetric vacuum spacetimes for $R^{1+\delta}$
gravity discussed in Ref.~\cite{Clifton:2005aj}.  We plan to
present more details in a forthcoming publication~\cite{us}.

\bigskip     

\begin{acknowledgments}
ALE and TLS acknowledge useful conversations with several
participants of session 86 of the Les Houches summer school. 
ALE acknowledges the support of an NSF Graduate Fellowship.  
MK and TLS are supported in part by DoE DE-FG03-92-ER40701, NASA
NNG05GF69G, and the Gordon and Betty Moore Foundation.  
\end{acknowledgments}

\end{document}